\documentclass[nofootinbib,preprintnumbers,superscriptaddress,notitlepage]{revtex4}
\usepackage{graphicx}

\begin{document}
\title{Evolution of Yukawa Couplings and Quark Flavour Mixings in the 5D MSSM}

\author{Ammar Abdalgabar}
\email[Email: ]{ammar.abdalgabar@students.wits.ac.za}
\affiliation{National Institute for Theoretical Physics; School of Physics, University of the Witwatersrand, Wits 2050, South Africa}

\author{A. S. Cornell}
\email[Email: ]{alan.cornell@wits.ac.za}
\affiliation{National Institute for Theoretical Physics; School of Physics, University of the Witwatersrand, Wits 2050, South Africa}

\begin{abstract}
The evolution equations of the Yukawa couplings and quark mixings are derived for the one-loop renormalization group equations in the 5D Minimal Supersymmetric Standard Model on an {$S^1 / Z_2$} orbifold. Different possibilities for the matter fields are discussed such as the cases of bulk propagating or brane localised fields. We discuss in both cases the evolution of the mass ratios and the implications for the mixing angles. 
\end{abstract}

\preprint{WITS-CTP-111}
\maketitle


\section{Introduction}

\par A theory of fermion masses and the associated mixing angles provide an interesting puzzle and a likely window to physics beyond the Standard Model (SM), where one of the main issues in particle physics is to understand the fermion mass hierarchy and mixings. A clear feature of the fermion mass spectrum is \cite{Falcone:2001ep} 
\begin{equation}
m_u \ll m_c \ll m_t \; , \; m_d \ll m_s \ll m_b \; , \; m_e \ll m_{\mu} \ll m_{\tau} \; .
\end{equation}
There have been many attempts to understand the fermion mass hierarchies and their mixings by making use of Renormalization Group Equations (RGEs) especially for Universal Extra Dimension (UED) models and their possible extensions (see Refs. \cite{Cornell:2010sz, Cornell:2011fw} and references therein). There are several versions of UED models, the simplest being the case of one flat extra dimension compactified on an $S^1/Z_2$ orbifold which has a size $1/R \approx 1$ TeV. This compactification will lead to a tower of new particle states in the effective 4-dimensional (4D) theory. As such, in the 4D effective theory there appears an infinite tower of massive Kaluza-Klein (KK) states, with a mass contribution inversely proportional to the radius of the extra-dimension. Due to the orbifolding mechanism the momentum is no longer conserved along the fifth dimension, and the symmetry is reduced to a KK parity conservation, $P = (-1)^n$ , which is now an exact symmetry, with $n$ referring to the KK level. That is, the UED models are effective theories in four dimensions with a cutoff $\Lambda$, with the consequence that the tree level spectrum is highly degenerate and where loop corrections to masses become important \cite{Cheng:2002iz}. The phenomenology of these UED models will arise when their flat extra dimensions allow all the SM fields (or some subset) to propagate in the full space-time \cite{Bhattacharyya:2009br, Liu:2011gr}.

\par There are many reasons to study such models, primarily as they provide a way to address the ``hierarchy problem", that is, the question of why the Planck scale $M_{Pl} \sim 10^{19}$ GeV, is so much smaller than the weak scale 246 GeV, but also to provide a means of breaking the electroweak symmetry, the generation of fermion mass hierarchies, and in studying the CKM matrix and new sources of CP violation \cite{Huang:2012kz, Barger:1992pk}. Furthermore, TeV scale grand unification and sources of dark matter are also possible in these theories \cite{Arkani, Dienes:1998vg}. To date much of the interest in UED models has been for its source of beyond the SM TeV-scale physics, largely arising from the tower of KK states approximately degenerate in mass at the scale set by the inverse of the compactification radius. KK parity and the 4D conservation of momentum imply that contributions to SM particle masses occur only for interactions at loop level, and that the lightest KK particle will be stable and a suitable dark matter candidate \cite{Rubakov}.

\par Recall that in extra-dimensional scenarios we are lead to a power law running of the gauge couplings due to the large number of KK states. These KK modes contribute the same way as the zeroth mode does, which we identify as the usual 4D Minimal Supersymmetric SM (MSSM) particles. In 5-dimensions, though, only Dirac fermions are allowed by the Lorentz algebra. As such there are eight supercharges which corresponds to the 4D view point of an $\mathcal{N} =2$ supersymmetry, where the $S^1 /Z_2$ orbifolding present in our UED models \cite{Cornell:2011fw} will break the $\mathcal{N} =1$ supersymmetry, for  more details  see Refs. \cite{Deandrea:2006mh, Arkani}.


\section{The 5D MSSM models}
 
\par In the 5-dimensional (5D) MSSM, the Higgs superfields and gauge superfields always propagate into the fifth dimension. However, different possibilities of localisation for the matter superfields can be studied. We shall consider the two limiting cases of superfields with SM matter fields all in the bulk or all superfields containing SM matter fields restricted to the brane. When all fields propagate in the bulk, the action for the matter fields $\Phi_i$ is \cite{Deandrea:2006mh, Cornell:2011fw}:
 \begin{eqnarray}
S_{matter} &=& \int d^8 z d y \left[ \bar{\Phi}_i \Phi_i+ \Phi_i^c \bar{\Phi}_i^c+ \Phi_i^c \partial_5 \Phi_i \delta(\bar{\theta})-\bar{\Phi}_i \partial_5 \bar{\Phi }_i^c \delta(\theta) \right. \nonumber \\
&& \left. \hspace{1cm} + \tilde{g} (2 \bar{\Phi}_i V \Phi_i - 2 \Phi_i^c  V \bar{\Phi}_i^c+ \Phi_i^c \chi \Phi_i \delta(\bar{\theta})+\bar{\Phi}_i \bar{\chi} \bar{\Phi }_i^c \delta(\theta) ) \right] \; .
\end{eqnarray}
Similarly, when all superfields containing SM fermions are restricted to the brane, the part of the action involving only gauge and Higgs fields is not modified, whereas the action for the superfields containing the SM fermions becomes:
\begin{equation}
 S_{matter} = \int d^8 z d y \delta(y) \left[ \bar{\Phi}_i \Phi_i + 2 \tilde{g}\bar{\Phi}_i V \Phi_i \right]\; . 
\end{equation}
For $\mathcal{N}= 1$, 4D supersymmetry the superfield formalism is well established: superfields describe quantum fields and their superpartners, as well as auxiliary fields, as a single object. This simplifies the notation and the calculation considerably. A similar formulation for a 5D vector superfield  and the superfield formulation for matter supermultiplets has been developed in Refs.\cite{Deandrea:2006mh, Arkani}. Note that in our model the Yukawa couplings in the bulk are forbidden by the 5D $\mathcal{N}= 1$ supersymmetry. However, they can be introduced on the branes, which are 4D subspaces with reduced supersymmetry. We will write the following interaction terms, called brane interactions, containing Yukawa-type couplings:
\begin{equation}
S_{brane}= \int d^8z dy \delta(y) \left(\frac{1}{6}\tilde{\lambda}_{{ijk}} \Phi_i \Phi_j \Phi_k \right) \delta(\bar{\theta})+ h.c. \; .
\end{equation}

\par In 4D MSSM, the one-loop correction to the gauge couplings are given by
\begin{equation}
16\pi^2 \frac{d g_i}{dt}=b_i g^3_i \; , \label{eqn:4DMSSMgauge}
\end{equation}
\noindent 
where $b_i=(\frac{33}{5}, 1, -3)$, $t=\ln(\mu/M_Z)$, and $M_Z$ is the $Z$ boson mass \cite{Cornell}.

\par In 5D MSSM, the one-loop corrections to gauge couplings are given by \cite{Cornell:2011fw, Bhattacharyya:2009br}
\begin{equation}
16\pi^2 \frac{d g_i}{d t}=(b_i +(S(t)-1) \tilde{b_i})g^3_i \; , \label{eqn:5DMSSMgauge}
\end{equation}
where $S(t)=e^t M_Z R$ is the number of KK states, {$\tilde{b_i}=(\frac{66}{5}, 10, 6)$} for matter fields in the bulk and {$\tilde{b_i}=(\frac{6}{5},-2,-6)$} for matter fields localised to the brane. As can be seen in Fig. \ref{fig1} (and Eq. (\ref{eqn:5DMSSMgauge})), the one-loop running of the gauge couplings changes with energy scale drastically and lowers the unification scale considerably. Specifically, for the compactification radii $R^{-1} = 1$, 4, 15 TeV, we find that the gauge couplings meet at around $\Lambda \approx$ 30, 120, 430 TeV respectively. As such the Yukawa couplings also receive finite corrections at each KK level whose magnitudes depend on the cutoff energy scale. The evolution of the Yukawa couplings were derived using the standard techniques of Refs. \cite{Deandrea:2006mh, Cornell:2011fw}. As such, the one-loop RGEs for Yukawa couplings in the 5D MSSM are: 
\begin{equation}
16 \pi^2 \frac{d Y_{i}}{d t} = Y_{i} \left[ T_{i} - G_{i}S(t) + F_{i} \right] \; , \label{eqn:YukawaRGE}
\end{equation}
where $i= u, d, e$ and the values of $G_i, F_i $ and $T_i$ are given in table \ref{tab1}. That is, when the energy scale $\mu > \frac{1}{R}$, or when the energy scale parameter $ t > \ln (\frac{1}{M_Z R})$ (where we have set $M_Z$ as the renormalization point) we shall use Eq.(\ref{eqn:YukawaRGE}), however, when the energy scale $M_Z < \mu < \frac{1}{R}$, the Yukawa evolution equations are dictated by the usual MSSM ones \cite{Cornell}.
 \begin{table}
\caption{\label{tab1}The terms present in the various Yukawa evolution equations, see Eq.(\ref{eqn:YukawaRGE}).}
\begin{center}
\begin{tabular}{cccc}
\hline
Scenarios & $G_u$ & $G_d$ & $G_e$  \\
\hline
Bulk & $\frac{13}{15}g_1^2+3g_2^2+\frac{16}{3}g_3^2$ & $\frac{7}{15}g_1^2+3g_2^2+\frac{16}{3}g_3^2$ & $\frac{9}{5}g_1^2+3g_2^2$  \\
\hline
Brane & $\frac{43}{30}g_1^2+\frac{9}{2}g_2^2+\frac{32}{3}g_3^2$ & $\frac{19}{30}g_1^2+\frac{9}{2}g_2^2+\frac{32}{3}g_3^2$ & $\frac{33}{10}g_1^2+\frac{9}{2}g_2^2$  \\
\hline
Scenarios & $T_d = T_e$ & $T_u$ & \\
\hline
Bulk & $(3 Tr ( Y_d^{\dagger} Y_d)+ Tr ( Y_e^{\dagger} Y_e))\pi S^2(t)$ & $3 Tr ( Y_u^{\dagger} Y_u)\pi S^2(t)$ & \\
\hline
Brane & $3 Tr( Y_d^{\dagger} Y_d)+ Tr( Y_e^{\dagger} Y_e)$ & $3 Tr( Y_u^{\dagger} Y_u)$ & \\
\hline
Scenarios  & $F_u$ & $F_d$ & $F_e$ \\
\hline
Bulk & $(3 Y_u^{\dagger} Y_u + Y_d^{\dagger} Y_d)\pi S^2(t)$ & $(3 Y_d^{\dagger} Y_d + Y_u^{\dagger} Y_u) \pi S^2(t)$ & $(3Y_e^{\dagger} Y_e)\pi S^2(t)$ \\
\hline
Brane & $(6Y_u^{\dagger} Y_u +2Y_d^{\dagger} Y_d) S(t)$ & $(6Y_d^{\dagger} Y_d +2Y_u^{\dagger} Y_u ) S(t)$ & $6Y_e^{\dagger} Y_e S(t)$ \\
\hline
\end{tabular}
\end{center}
\end{table}

\par Yukawa coupling matrices can be diagonalized by using two unitary matrices {$U$} and {$V$}, where 
 $$
 UY^{\dagger}_u Y_u U^{\dagger}= diag(f^2_u,f^2_c,f^2_t) \; , \; VY^{\dagger}_d Y_d V^{\dagger}= diag(h^2_d,h^2_s,h^2_b) \; .
 $$
The CKM matrix appears as a result (upon this diagonalisation of quark mass matrices) of {$V_{CKM} = U V^{\dagger}$}. The variation of the CKM matrix and its evolution equation for all matter fields in the bulk is \cite{Babu:1987im, Cornell:2010sz, Liu:2011gr}:
\begin{equation}
16 \pi^2 \frac{dV_{i \alpha}}{dt}= \pi S^2 \left[\sum_{\beta, j \neq i}{ \frac{f_i^2+f_j^2}{f_i^2-f_j^2} h_{\beta}^2 V_{i\beta} V^\ast_{j \beta}}V_{j \alpha}+\sum_{j, \beta\neq \alpha}{\frac{h_{\alpha}^2+h_{\beta}^2}{h_{\alpha}^2-h_{\beta}^2} f_j^2 V^\ast_{j\beta} V_{j \alpha}V_{i \beta}}\right]\; . \label{eqn:CKMRGE}
\end{equation}
For all matter fields on the brane, the CKM evolution is the same as Eq.(\ref{eqn:CKMRGE}) but  $\pi S^2$ is replaced by 2 $S$.

\par In making use of these equations for this current work, we recall that in the 4D MSSM the particle spectrum contains two Higgs doublets and the supersymmetric partners to the SM fields. After the spontaneous symmetry breaking of the electroweak symmetry, five physical Higgs particles are left in the spectrum. The two Higgs doublets $H_u$ and $H_d$, with opposite hypercharges, are responsible for the generation of the up-type and down-type quarks masses respectively. The vacuum expectation values (vevs) of the neutral components of the two Higgs fields satisfy the relation $v^2_u+v^2_d= (\frac{246}{\sqrt{2}})^2 = (174 GeV)^2$. The fermions mass matrices appear after the spontaneous symmetry breaking from the fermion-Higgs-Yukawa couplings. As a result, the initial Yukawa couplings are given by the ratios of the fermion masses to the appropriate Higgs vev as follows:
\begin{equation}
f_{u,c,t} = \frac{m_{u,c,t}}{v_u}\; , \; h_{d,s,b} = \frac{m_{d,s,b}}{v_d}\; , \; y_{e,\mu,\tau} = \frac{m_{e,\mu,\tau}}{v_d}\; ,
\end{equation}
where we define {$tan \beta=\frac{v_u}{v_d}$}, which is the ratio of vevs of the two Higgs fields $H_u$ and $H_d$.


\section{Numerical Results and Conclusions}

\par For our numerical calculations we assume the fundamental scale is not far from the range of the LHC and set the compatification radii to be {$R^{-1}= 1$} TeV, 4 TeV and 15 TeV. Note that these values are chosen to highlight that once the first KK threshold is crossed we see the same phenomenological running to the cut-off scale. Only some selected plots will be shown and we will comment on the other similar cases not explicitly presented. We quantitatively analyse these quantities in the 5D MSSM for {$tan\beta $} small {$(tan\beta =5)$}, intermediate {$(tan\beta =30)$}, and large {$(tan\beta =50)$}, though we observed similar behaviours for all values of {$tan\beta $} (note that all plots are given for {$tan\beta =30$} for illustrative purposes only); the initial values we shall adopt at the $M_Z$ scale are: for the gauge couplings $\alpha_1$($M_Z$) = 0.01696, $\alpha_2$($M_Z$) = 0.03377, and $\alpha_3$($M_Z$) = 0.1184; for the fermion masses $m_u$($M_Z$) = 1.27 MeV, $m_c$($M_Z$) = 0.619 GeV, $m_t$($M_Z$) = 171.7 GeV, $m_d$($M_Z$) = 2.90 MeV, $m_s$($M_Z$) = 55 MeV, $m_b$($M_Z$) = 2.89 GeV, $m_e$($M_Z$) = 0.48657 MeV, $m_\mu$($M_Z$) = 102.718 MeV, and $m_\tau (M_Z) = 1746.24$ MeV \cite{Cornell:2011fw, Xing:2007fb}.

\begin{figure}
\begin{center}
\includegraphics[width=7cm,angle=0]{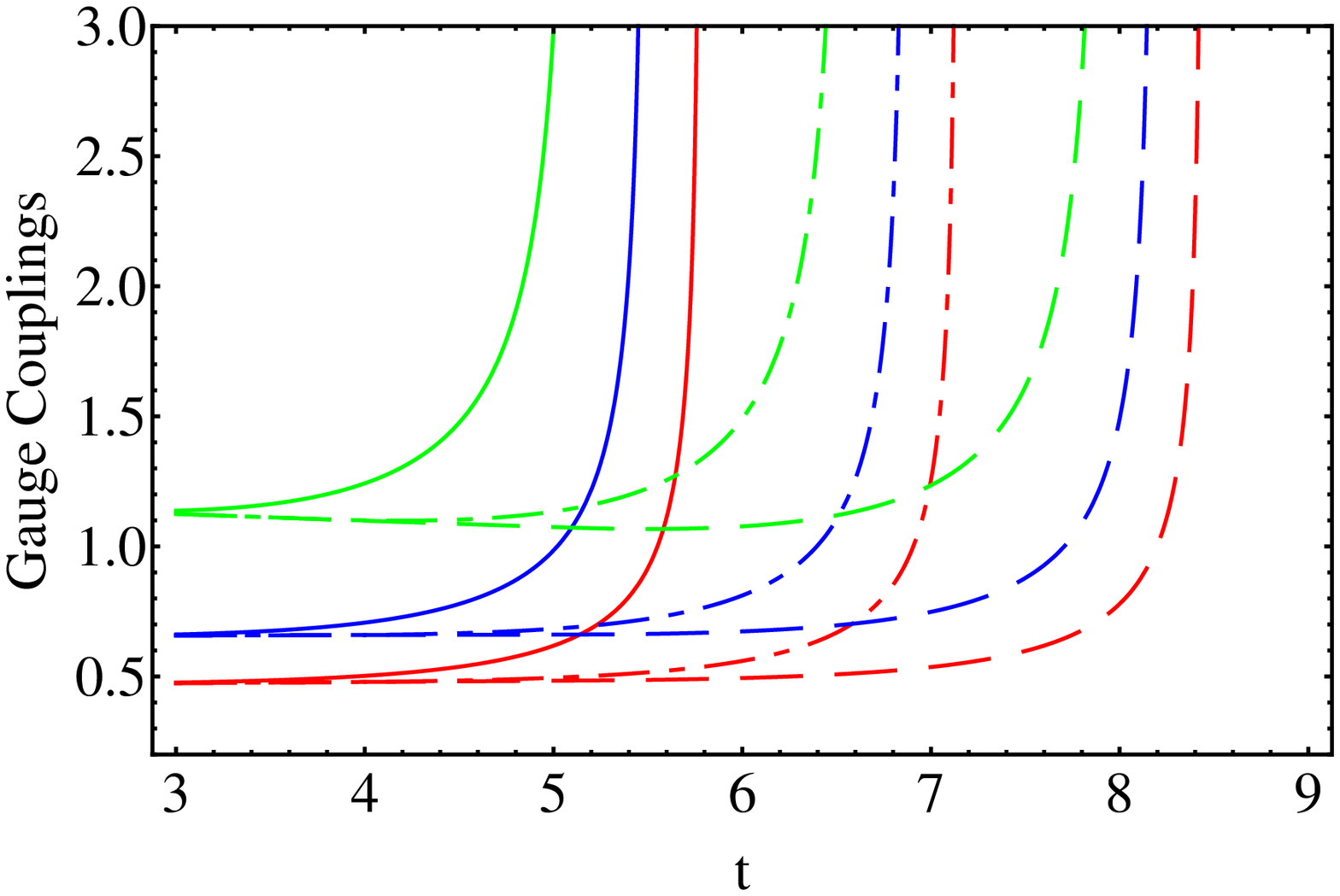} \qquad
\includegraphics[width=7cm,angle=0]{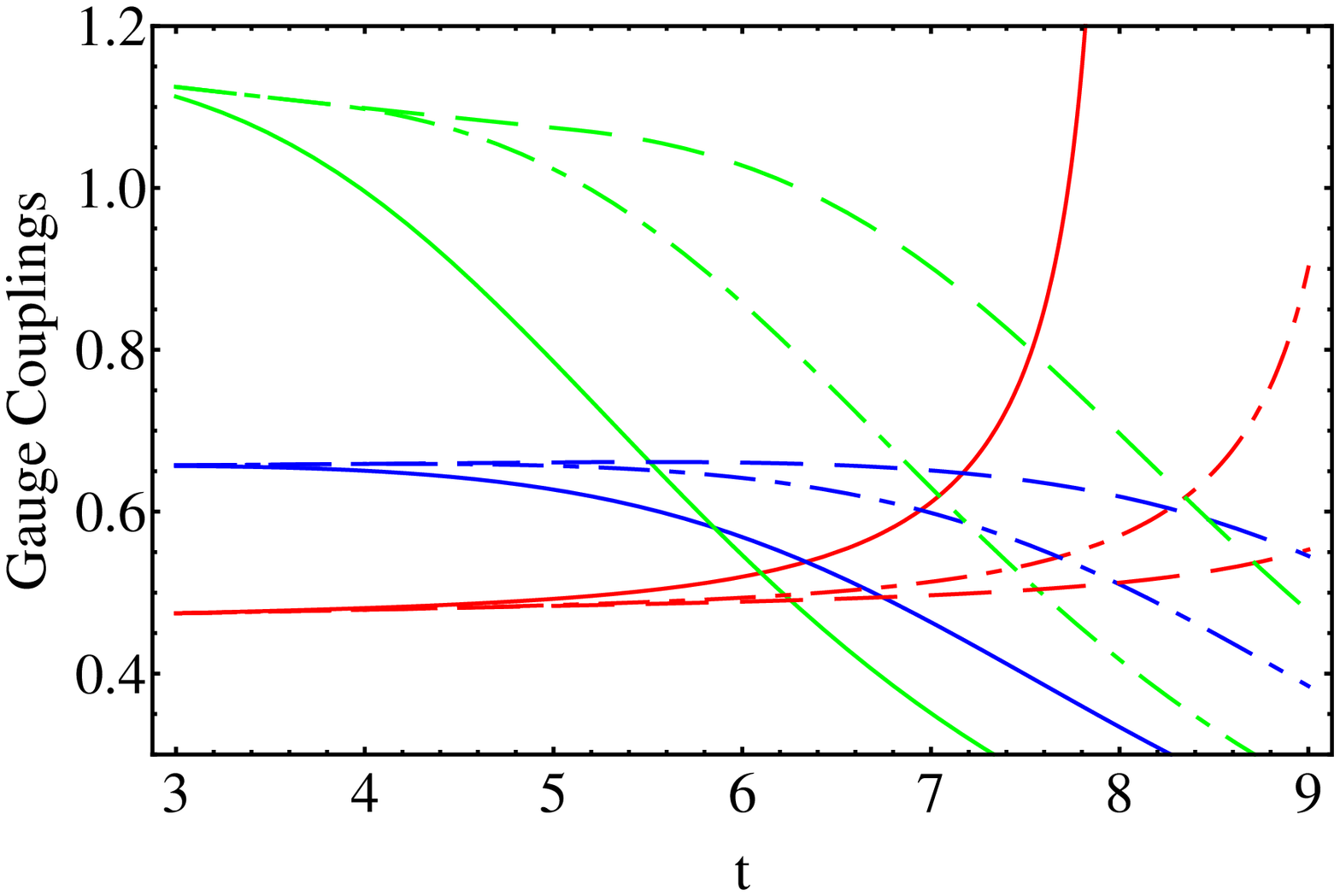}
\end{center}
\caption{\label{fig1}(Colour online) Gauge couplings ({$g_1$} (red), {$g_2$} (blue), {$g_3$} (green) with: in the left panel, all matter fields in the bulk; and the right panel for all matter fields on the brane; for three different values of the compactification scales (1 TeV (solid line), 4 TeV (dot-dashed line), 15 TeV (dashed line))) as a function of the scale parameter {$t$}.}
\end{figure}
\begin{figure}
\begin{center}
\includegraphics[width=7.5cm,angle=0]{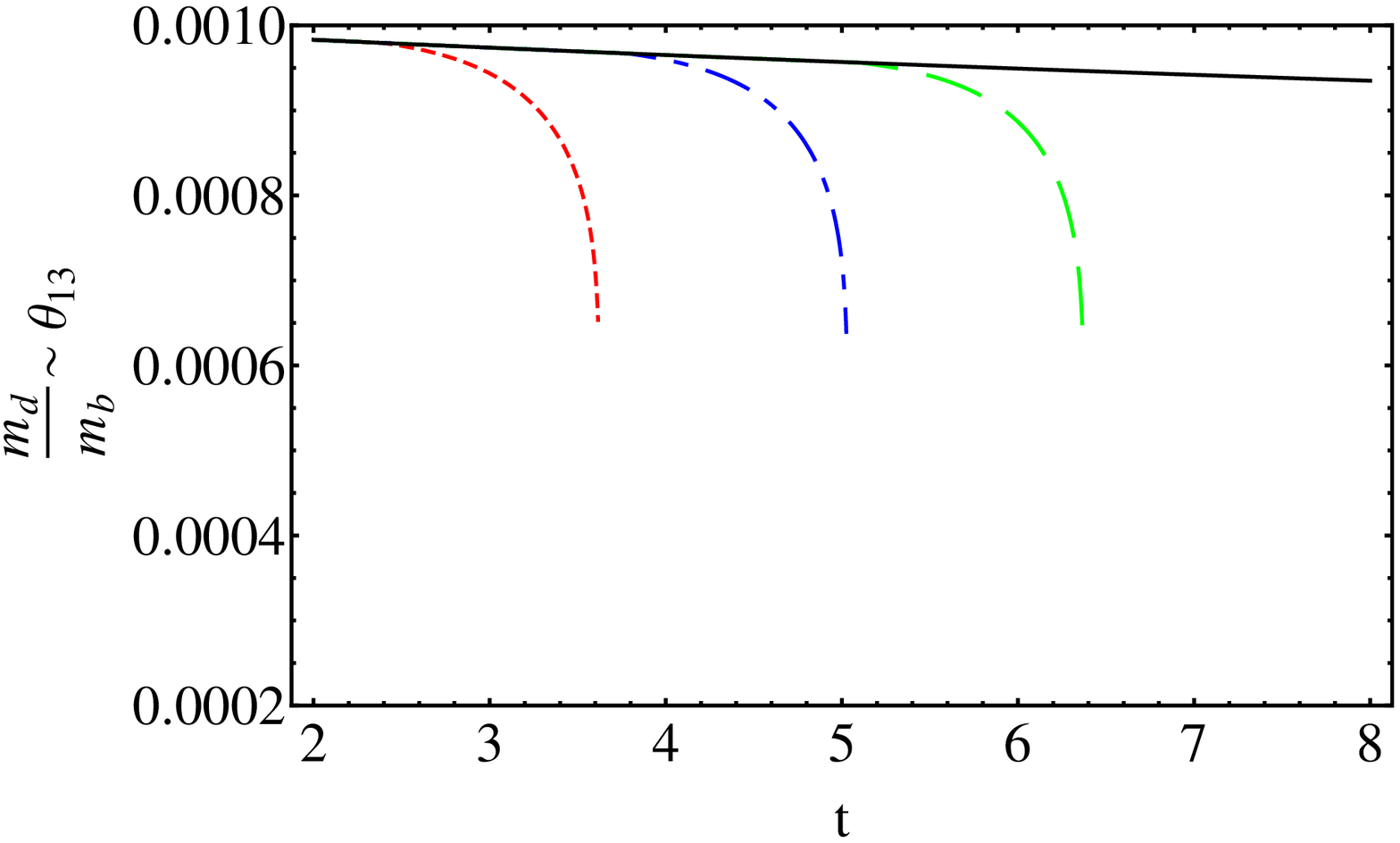} \qquad
\includegraphics[width=7.5cm,angle=0]{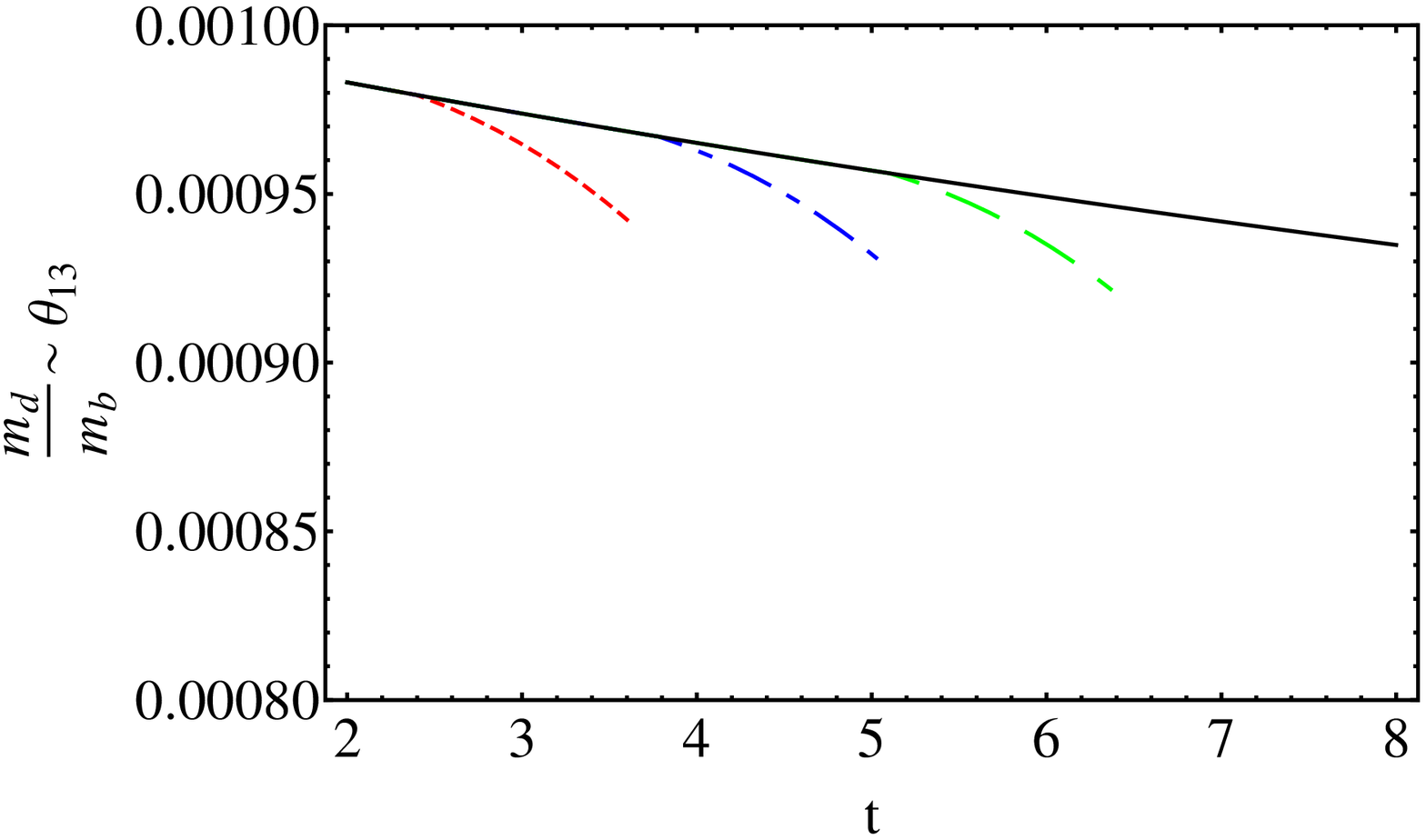}
\end{center}
\caption{\label{fig2}(Colour online) Evolution of the mass ratio {$\frac{m_d}{m_b}$}, with: in the left panel, all matter fields in the bulk; and the right panel for all matter fields on the brane. Three different values of the compactification radius have been used 1 TeV (dotted red line), 4 TeV (dot-dashed blue line), 15 TeV (dashed green line), all as a function of the scale parameter {$t$}.}
\end{figure}
\begin{figure}
\begin{center}
\includegraphics[width=7.5cm,angle=0]{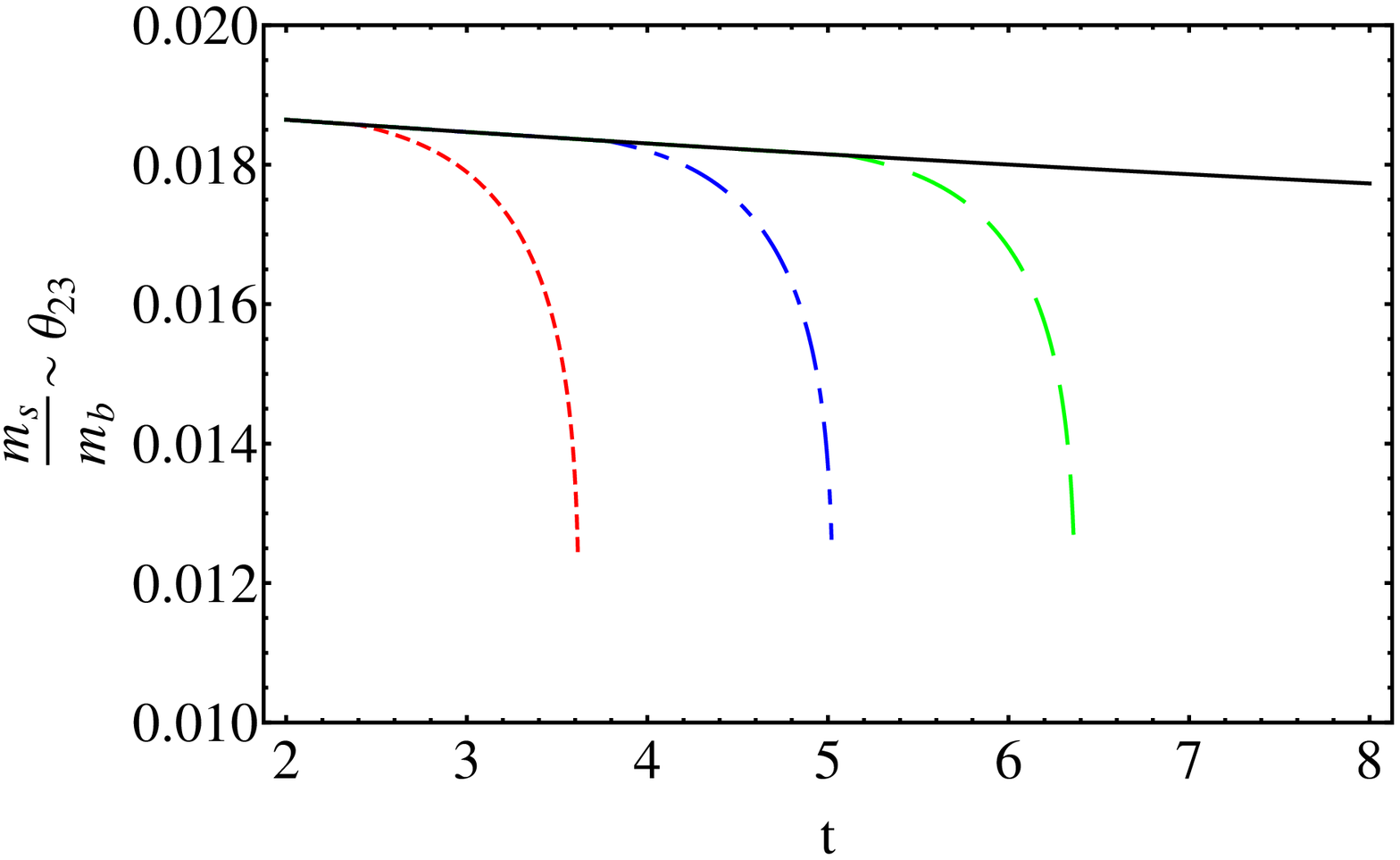} \qquad
\includegraphics[width=7.5cm,angle=0]{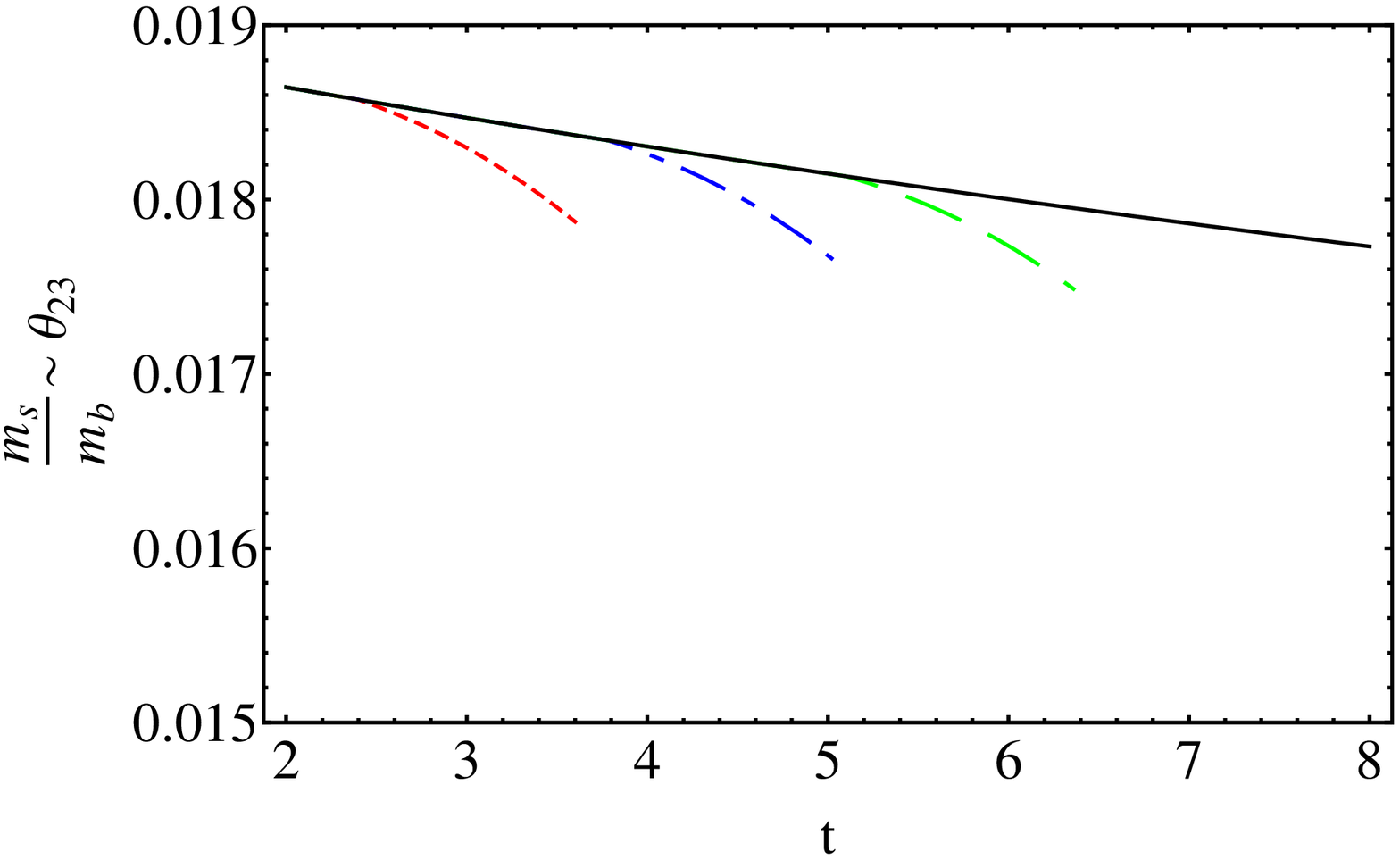}
\end{center}
\caption{\label{fig3}(Colour online) Evolution of mass ratio {$\frac{m_s}{m_b}$}, with the same notations as Fig.\ref{fig2}.}
\end{figure}
\begin{figure}
\begin{center}
\includegraphics[width=7.5cm,angle=0]{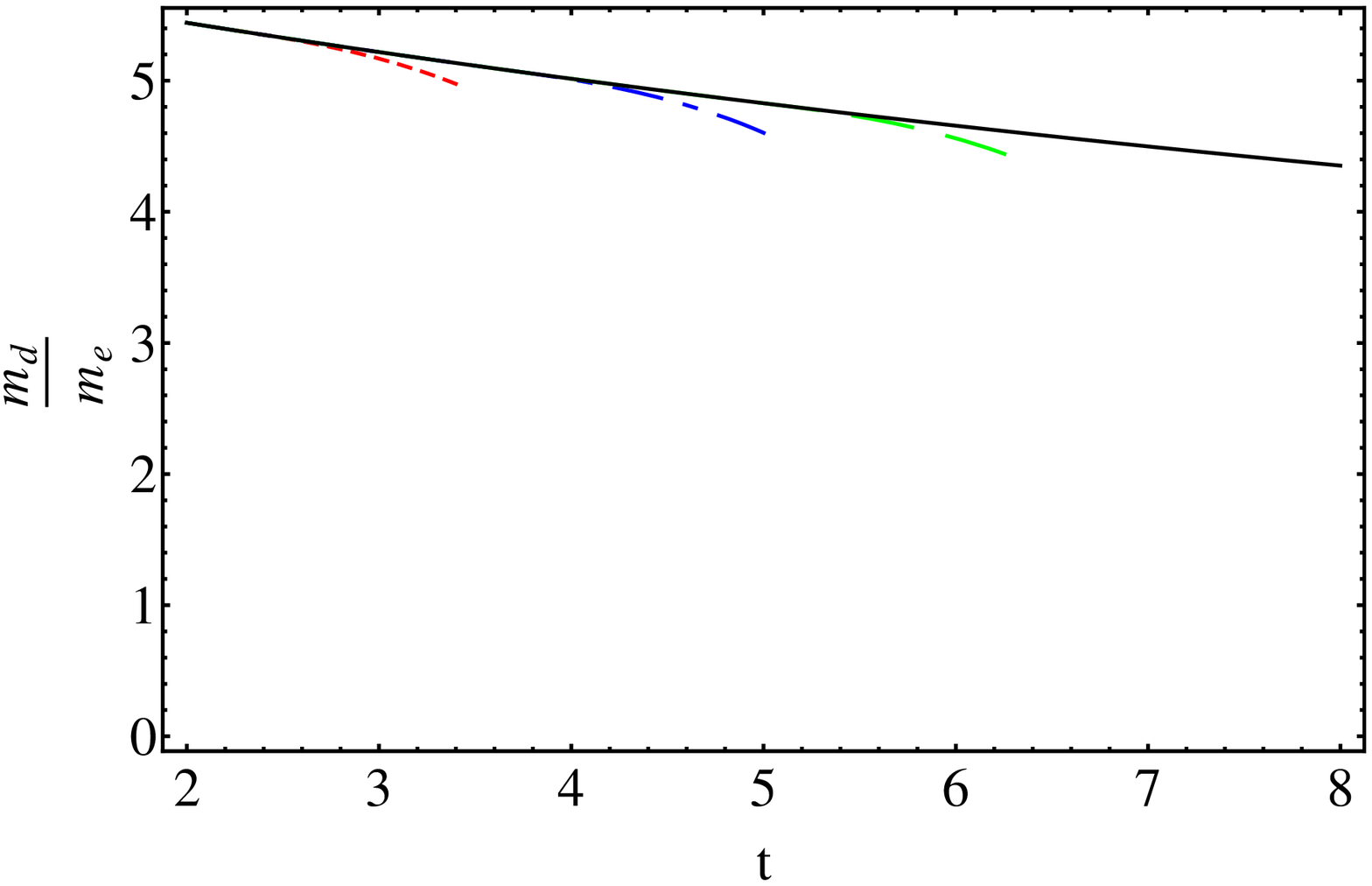} \qquad
\includegraphics[width=7.5cm,angle=0]{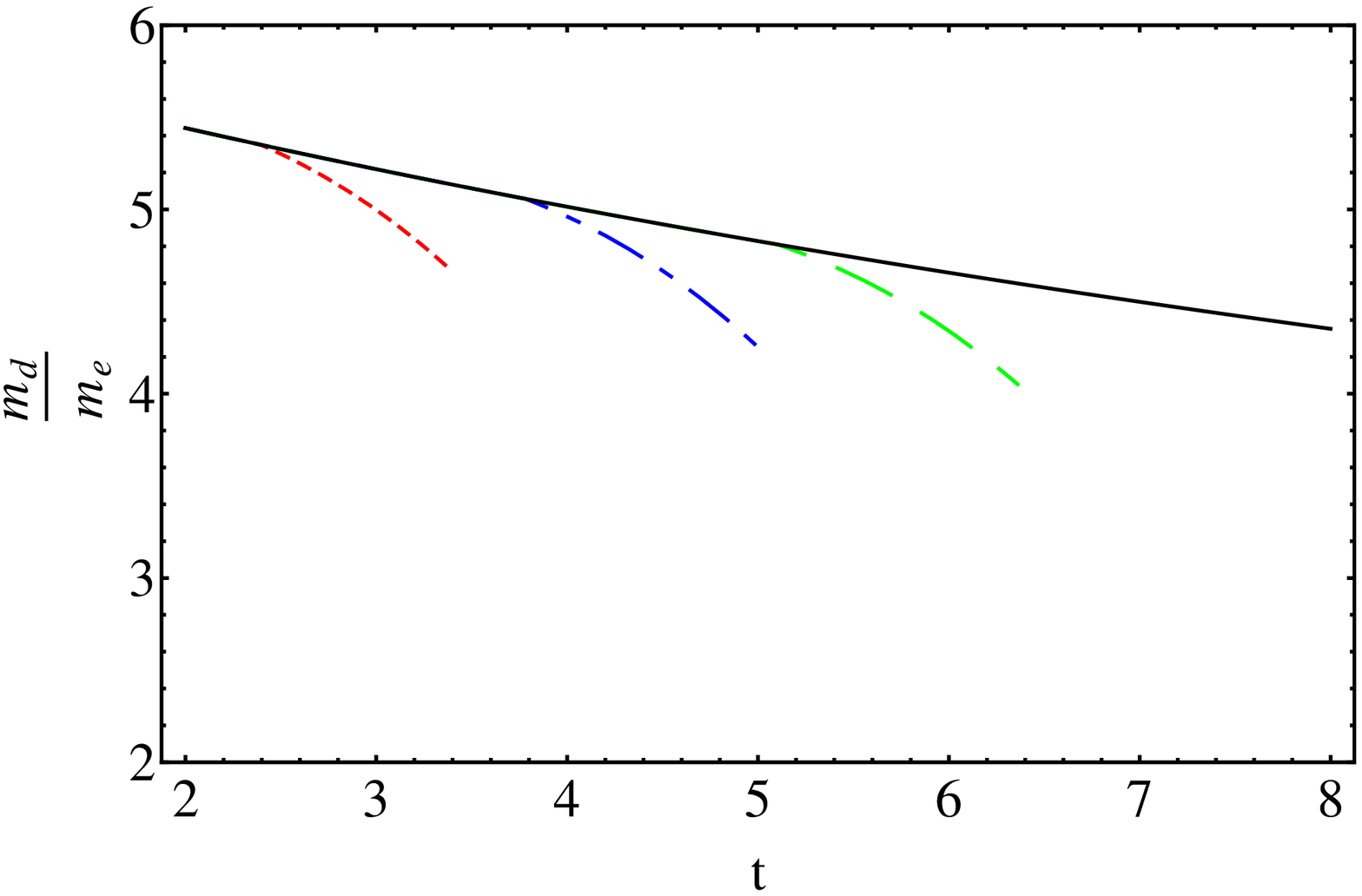}
\end{center}
\caption{\label{fig4}(Colour online) Evolution of mass ratio {$\frac{m_d}{m_e}$}, with the same notations as Fig.\ref{fig2}.}
\end{figure}
\begin{figure}
\begin{center}
\includegraphics[width=7.5cm,angle=0]{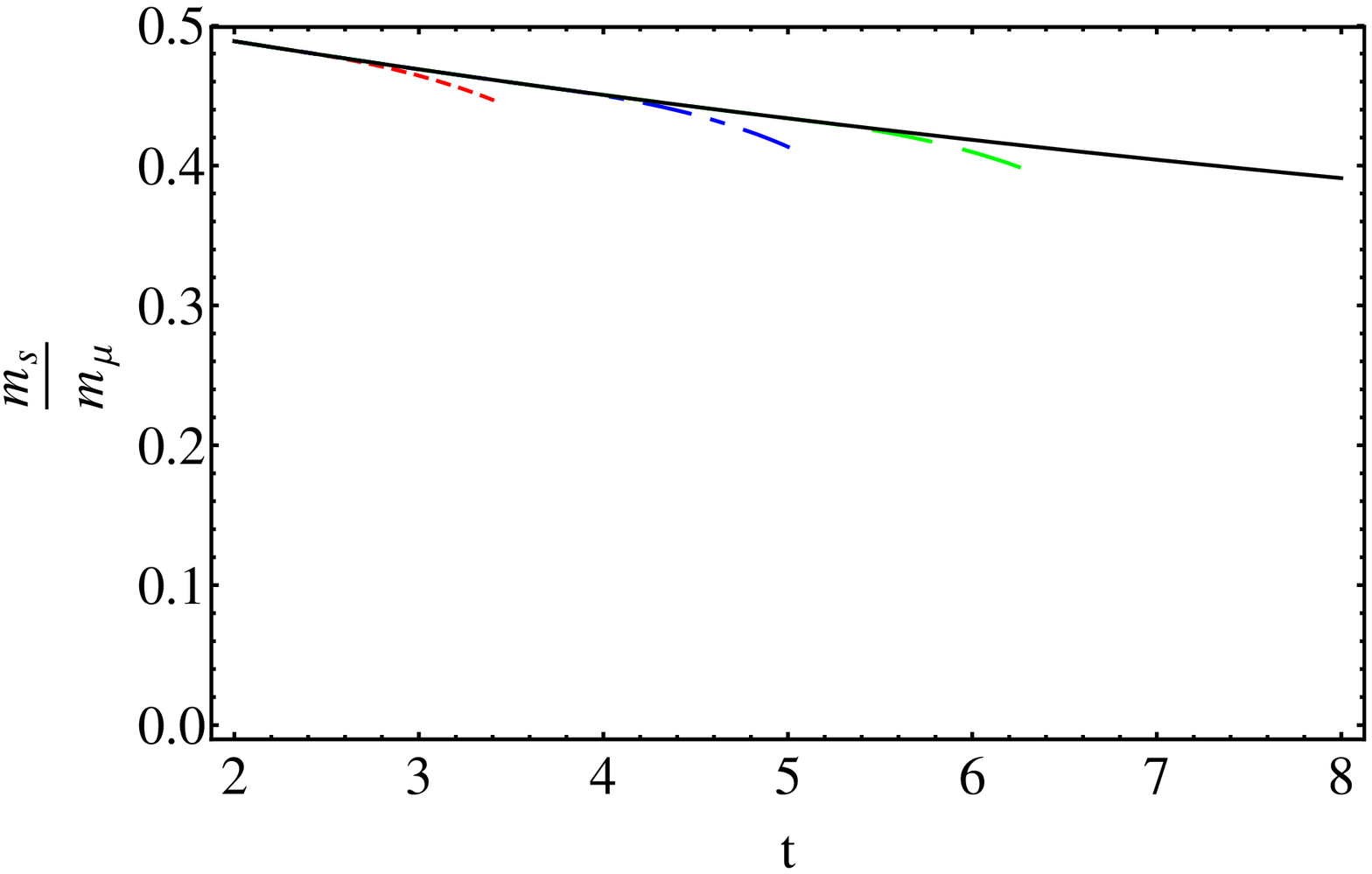} \qquad
\includegraphics[width=7.5cm,angle=0]{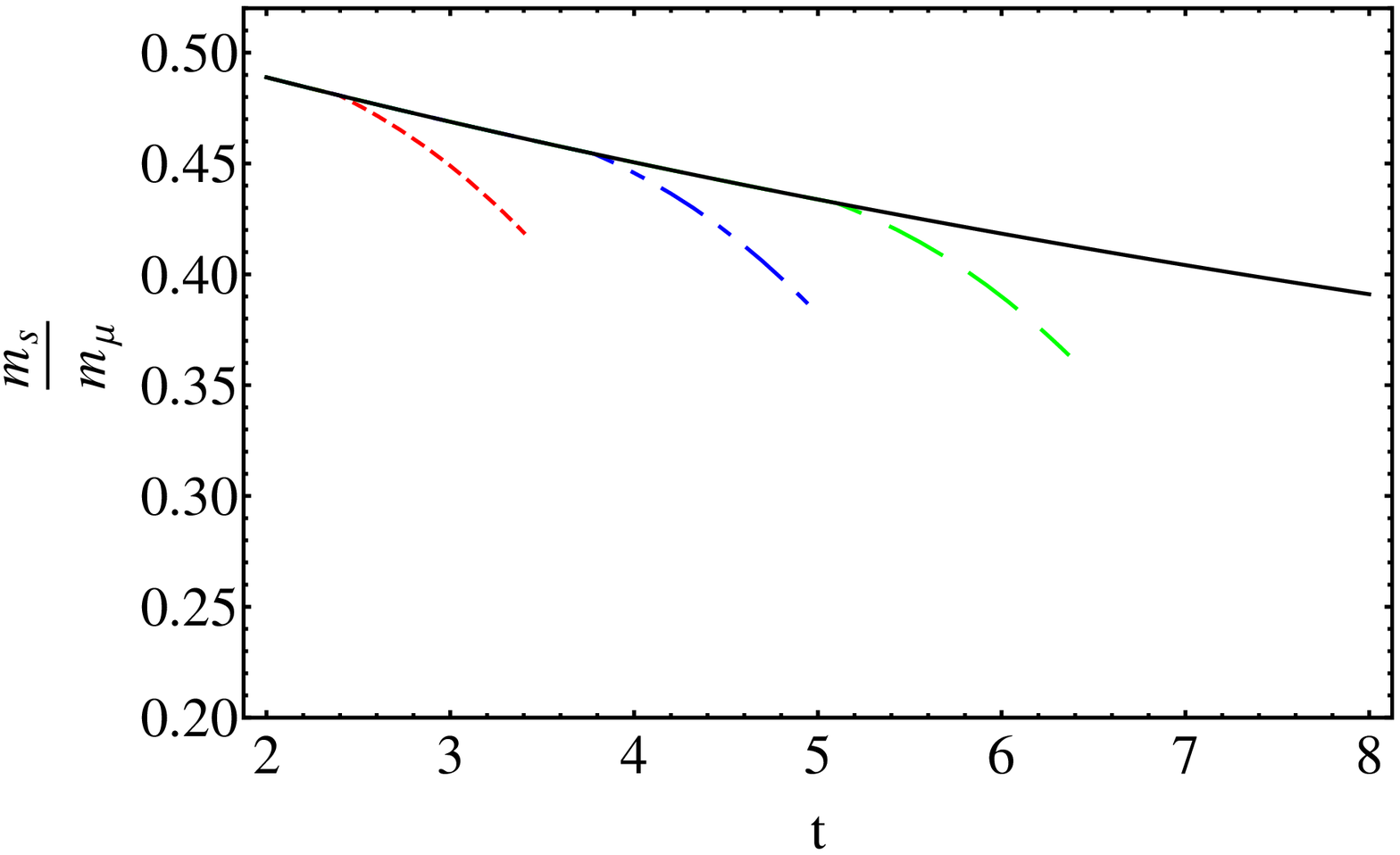}
\end{center}
\caption{\label{fig5}(Colour online) Evolution of mass ratio {$\frac{m_s}{m_{\mu}}$}, with the same notations as Fig.\ref{fig2}.}
\end{figure}
\begin{figure}
\begin{center}
\includegraphics[width=7.5cm,angle=0]{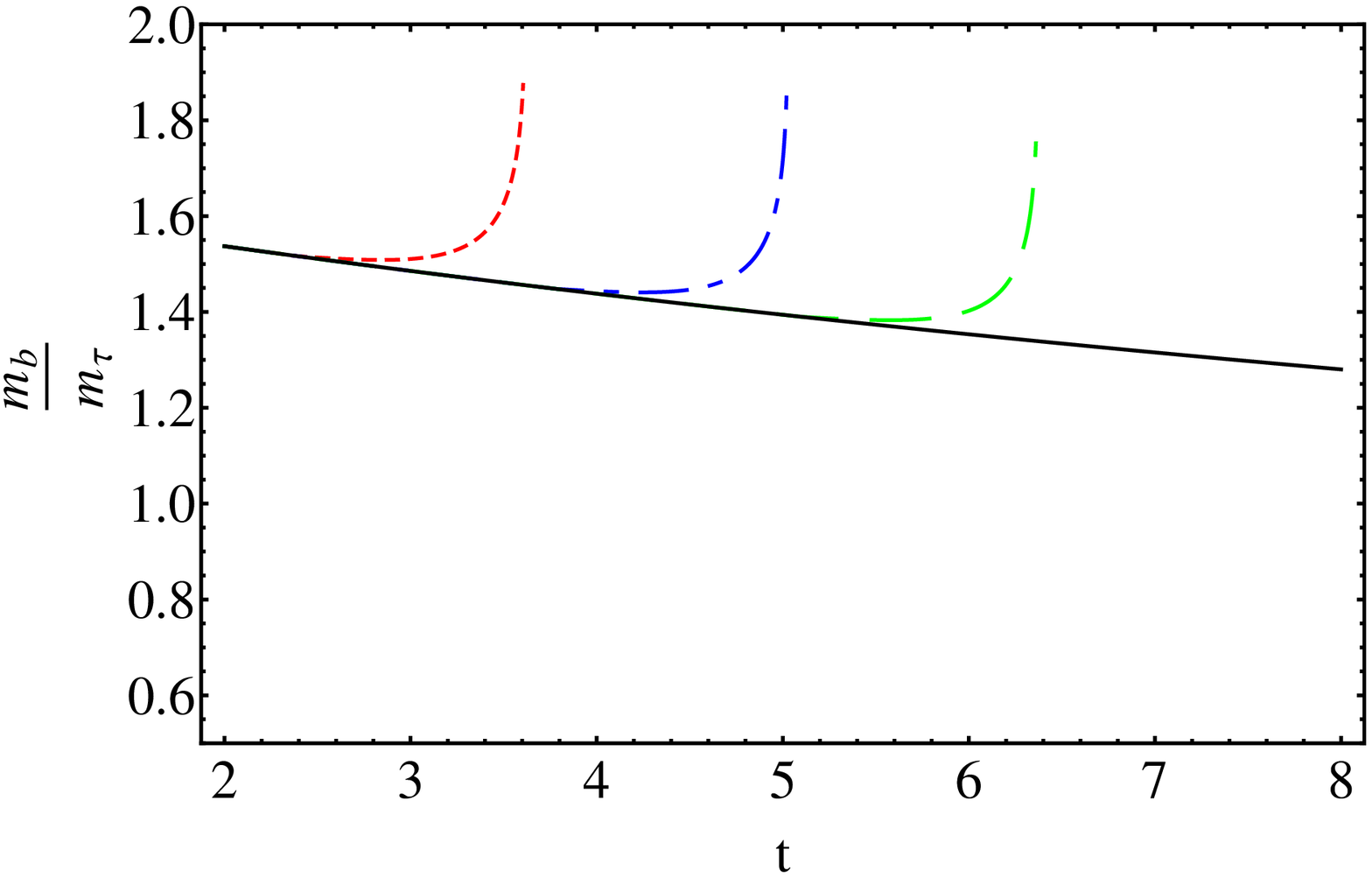} \qquad
\includegraphics[width=7.5cm,angle=0]{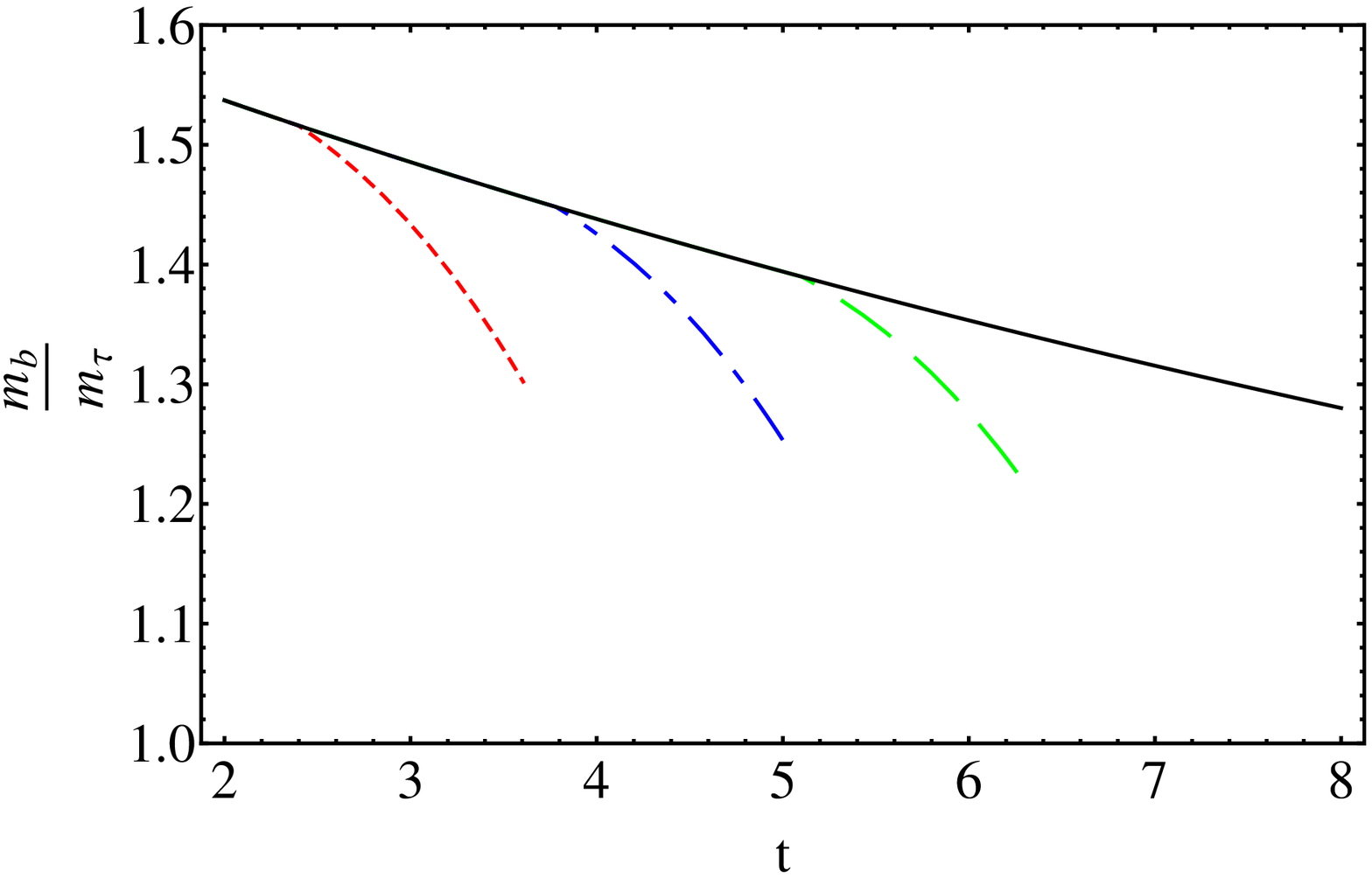}
\end{center}
\caption{\label{fig6}(Colour online) Evolution of mass ratio {$\frac{m_b}{m_{\tau}}$}, with the same notations as Fig.\ref{fig2}.}
\end{figure}

\par As illustrated in Figs. \ref{fig2} and \ref{fig3} the mass ratios evolve in the usual logarithmic fashion when the energies are below 1 TeV, 4 TeV and 15 TeV respectively. However, once the first KK threshold is reached the contributions from the KK states become increasingly significant and the effective 4D MSSM couplings begin to deviate from their normal trajectories. They evolve faster and faster after that point, their evolution  diverging due to the faster running of the gauge couplings, where in approaching our cutoff for the effective theory, $\Lambda$, any new physics would then come into play, see Fig.\ref{fig1}. As such, we have chosen cutoffs for our effective theory of; for the bulk case, where $g_3$ becomes large (and perturbation theory breaks down), and for the brane case, where $g_2 = g_3$ (and an expected mechanism for unification would take over). As such, the one-loop running of the gauge couplings changes with energy scale drastically and lowers the unification scale considerably. Specifically, for the compactification radii $R^{-1} = 1$, 4, 15 TeV, we find that for the brane localised matter fields case the cut-off $\Lambda \approx$ 30, 120, 430 TeV respectively. For the bulk case the cut-off is around $\Lambda \approx$ 6, 30, 70 TeV respectively.

\par On the other hand, in $SU(5)$ theory we have {$m_d = m_e$}, {$m_s = m_{\mu}$} and {$m_b = m_{\tau}$} at the unification scale, where in 5D MSSM, due to power law running of the Yukawa couplings, the renormalization effects on these relations can be large for {$\frac{m_d}{m_e}$} and {$\frac{m_s}{m_{\mu}}$}, for both scenarios see Figs. \ref{fig4} and \ref{fig5}. We have shown, by numerical analysis of the one loop calculation, that the mass ratios {$\frac{m_d}{m_e}$}, {$\frac{m_s}{m_{\mu}}$} and {$\frac{m_b}{m_{\tau}}$} decrease as energy increases. However, {$\frac{m_b}{m_{\tau}}$} for  matter fields in the bulk increases as energy increases see Fig. \ref{fig6} left panel.

\par As depicted in Fig. \ref{fig6}, for the third generation the mass ratios increase rapidly as one crosses the KK threshold at {$\mu= R^{-1}$} for the bulk case, resulting in a rapid approach to a singularity before the unification scale is reached. However, for the brane localised case the contribution from the gauge couplings may become significant, therefore the trajectory might change direction, and the mass ratios decrease instead of increasing. 

\par In conclusion, in this paper the mass ratios in 5D MSSM on a {$S^1 / Z_2$} orbifold, for different possibilities of matter field localisation, were discussed. That is, where they are either bulk propagating or localised to the brane. We found that the 5D MSSM has  substantial effects on the scaling of fermion masses for both cases, including both quark and lepton sectors. We quantitatively analysed these quantities in the 5D MSSM with small, intermediate, and large {\textit{$tan\beta $}} values, though we observed similar behaviours for all values of {\textit{$tan\beta $}}. We have shown that the scale dependence is no longer logarithmic, having a power law behaviour. We also found that for both scenarios the theory is valid up to the unification scale receiving significant renormalization group corrections. Therefore the 5D MSSM promises exciting phenomenology for upcoming collider physics results. 


\end{document}